\newcommand{\be}{\begin{eqnarray}}
\newcommand{\ee}{\end{eqnarray}}
\newcommand{\nn}{\nonumber \\}
\newcommand{\V}{{{\cal V}}}

\newcommand{\A}{{{\cal A}}}

\newcommand{\openone}{\mbox{1\kern -0.25em I}}
\newcommand{\openK}{\mbox{I\kern -0.25em K}}
\newcommand{\openZ}{\mbox{Z\kern -0.4em Z}}
\newcommand{\openR}{\mbox{I\kern -0.25em R}}
\newcommand{\openH}{\mbox{I\kern -0.25em H}}
\newcommand{\openM}{\mbox{I\kern -0.25em M}}
\newcommand{\openC}{\mbox{C\kern -0.55em I\hspace{0.25em}}}
\newcommand{\pprime}{\prime\prime}
\newcommand{\Id}{\mbox{I\kern -0.1em d}}

\newcommand{\dwedge}{{\,\dot{\wedge}\,}}

\documentstyle[eqsecnum,aps]{revtex}

\begin{document}
%\draft
\preprint{Tue-prep-1101}
\title{Vertex functions and generalized normal-ordering by
triple systems in non-linear spinor field models%
\footnote{This article is based on chapter two of the author's
thesis \cite{Faus-thesis}}
}
\author{Bertfried Fauser}
\address{Eberhard-Karls-Universit\"{a}t\\
Institut f\"ur Theoretische Physik\\
Auf der Morgenstelle 14\\
72072 T\"ubingen {\bf Germany}\\
Electronic mail: Bertfried.Fauser@uni-tuebingen.de
}
%\date{\today}
\date{November 8, 1996}
\maketitle
\begin{abstract}
Triple systems are closely related to Yang-Baxter symmetries.
Utilizing a non-parameter-dependent triple product, we derive
the BCS interaction. The enlargement of the notion of symmetry
leads in some sense to a regular vertex function. The connection
to the effect of running coupling constants is outlined, which
leads to the recently discussed anisotropic effective local
interactions. Furthermore, a discussion of the physical nature
of q-symmetries is given. 
\end{abstract}

\section{\protect\label{Sec-1}Introduction}

In a couple of papers, Susumu Okubo has shown the equivalence of
triple products and the Yang-Baxter symmetry
\cite{Okub1,Okub2,Okub3}. These triple
systems\cite{Triple1,Triple2,Triple3} provide an easy
way to find solutions of the Yang-Baxter equation in several
cases. Okubo distinguishes orthogonal triple systems (OTS) and
symplectic triple systems (STS) due to the symmetry of the
involved bilinear forms. 

In this short note, we want to emphasize a new direction of the
application of this method. The main idea is developed from the
property of triple products to map $\V \otimes \V \otimes \V
\mapsto \V$. Since an anharmonic interaction is of such a cubic
structure, we are able to utilize this structure for a
simplification. The same argument holds for the vertex function
of non-linear spinor fields.

{}From a historic point, triple systems are related to
non-associative, alternative algebras and especially Jordan
algebras. These algebras were developed in the same time as the
von Neumann quantum mechanics \cite{vonNeu,Dirac} to avoid the
complex numbers \cite{Jord33,Jord34}. This was expressed in the
name $r$-number contrary to $c$-numbers or $q$-numbers of the
new mechanics. Already Jordan proofed, that the Jordan algebras
are equally well suited to handle QM as the operator algebras
are, except from some exceptional cases sic. the exceptional
Jordan algebras. Due to the difficulties in calculating with 
non-associative algebras, Jordan algebras were not accepted in
physics beside some special cases. 

So they became useful in supersymmetric string models and
conformal field theory \cite{Guena91,Guena92}. The Nambu
mechanics \cite{Namb73}, which is a generalization of Hamilton
mechanics by multiple Hamiltonians leads to Jordan algebra
structures. The quantum analog is related to para-fermions
\cite{Deberg,Beckers} as was shown by Yamaleev and others
\cite{Yama-mex,Yama-aach}. In this context, it is important to
note, that Jordan algebras appear naturally as subalgebras of
Clifford algebras \cite{Sobc81}, which are closely related to
the spinor fields we want to employ in section \ref{SEC-3}.
Furthermore, we find a close connection between so called
generalized Clifford algebras \cite{Dixo78,Dixo81,Rama85} and
the Duffin-Kemmer-Petiau equations and matrices
\cite{Roma61,Krol93b}, as again to para-fermions. This motivates
the attempt to employ triple systems and triple products to
spinor modules. 

The relation between triple systems and non-associative algebras
is established via the notation of the associator, which can be
defined in an algebra $\A$ with the binary, linear relations
($a,b,c,c^\prime \in \A$)
\be
+ &:& \A \times \A \mapsto \A \nn
  &&  (a;b) \rightarrow a+b=b+a=c \nn
m &:& \A \times \A \mapsto \A \nn
  &&  (a;b) \rightarrow m(a,b)=ab=c^\prime,
\ee
subjected to the usual compatibility requirement of the
distributive law
\be
m(a,b+c)&=&m(a,b)+m(a,c) \nn
m(b+c,a)&=&m(b,a)+m(c,a)
\ee
as the ternary relation -- associator:
\be
(,,) &:& \A \times \A \times \A \mapsto \A \nn
     &&  (a;b;c) \rightarrow (a,b,c) = (ab)c-a(bc).
\ee
The associator is thus a measurement of the deviation from
associativity in an analogous sense as the commutator is a
measure for the lack of commutativity.

Furthermore, we want to emphasize, that ternary algebraic
structures are of recent interest for its own \cite{Kerner1},
because one may describe quark properties very elegant within
them. Also one has to notice, that not every ternary
multiplication relation can be expressed as an algebraic
expression of the binary multiplication and addition
\cite{Kerner2}, as we do in this paper.

The paper is organized as follows:

In section \ref{SEC-2} we give the definition of OTS. Thereby we
state in \ref{Sec-2a} the defining relations in an axiomatic
way, establish in \ref{Sec-2b} the connection to the
Yang-Baxter-equation and provide in \ref{Sec-2c} a motivation of
the defining relations. Their physical relevance is emphasized. 

Section \ref{SEC-3} is devoted to the application of a
particular triple system and a discussion of its generalization
and physical implications. A vertex function of a non-linear
spinor field model is obtained due to strict application of
triple product methods in \ref{Sec-3a}, which in the particular
case results in the BCS dynamics when restricted to the low
energy -- non-relativistic -- regime.

Subsection \ref{Sec-3b} describes a generalization of the
results in \ref{Sec-3a} and connects the transition from
non-$\theta$-dependent triple systems to $\theta$-dependent ones
with a generalized ordering of interacting quantum theories. The
connection to normal-ordering -- for the special case discussed in
\ref{Sec-3a} -- as the relevance to recent developements is given.

Subsection \ref{Sec-3c} speculates about an alternative
interpretation of $q$-deformed variables i.e. the quantum
(hyper) plane and $q$-symmetry. The results from \ref{Sec-3a}
and \ref{Sec-3b} motivate and support the new approach to look
at $q$-symmetry as the symmetry of composite structures and {\it
not}\/ as a fundamental physical issue. 

The paper closes with a summary in section \ref{SEC-4}.

\section{\protect\label{Sec-2}Orthogonal triple systems}

In this section we give an outline of the mathematical structure
of triple systems. We follow closely Okubo
\cite{Okub1,Okub2,Okub3}. Because of the aim to apply these
techniques to non-linear spinor fields, most emphasis is laid
in this preparatory section on a physical motivation of triple systems.

\subsection{\label{Sec-2a}Definition of orthogonal triple systems}

Let $\V$ be a finite dimensional $\openK$-vector space of
dimension $N$ over $\openK$. We denote elements of $\V$ with
small latin letters. Let $\V$ be equipped with a non-degenerate
bilinear form $<.\vert .>$ -- scalar product --, hence with
$x,y\in \V$ $\alpha\in \openK$
\be
<.\vert .> &:& \V \times \V \mapsto \openK \nn
           &&  <x\vert y>=\alpha \nn
           &&  <x\vert y>=0\quad\forall y \Rightarrow x\equiv 0.
\ee
Denote the triple product as $[x,y,z]$ with range in $\V$. An
orthogonal triple system is defined by the following conditions,
where $x,y,z,u,v,w\in \V$~and $\alpha,\beta\in \openK$
\be
\label{triple-def}
i)  &~~& <x\vert y>=<y\vert x>\nn
ii) &~~& [x,y,z]=[y,z,x]=-[z,y,x]\nn
iii)&~~& <w\vert [x,y,z]>\quad\mbox{antisymmetric in
       $x,y,z$ and $w$}\nn
iv) &~~& <[x,y,z]\vert [u,v,w]>=\alpha\Big\{\nn
&~~& <x \vert u><y \vert v><z \vert w>+<y \vert u><z \vert v><x \vert w>\nn
&~~&+<z \vert u><x \vert v><y \vert w>-<x \vert w><y \vert v><z \vert u>\nn
&~~&-<y \vert w><z \vert v><w \vert u>-<z \vert w><x \vert v><y \vert v>
\Big\}\nn
&~~&
+\beta\Big\{<x \vert u><y \vert [z,v,w]>+<y \vert u><z \vert [x,v,w]>\nn
&~~&+<z \vert u><x \vert [y,v,w]>+<x \vert v><y \vert [z,w,u]>\nn
&~~&+<y \vert v><z \vert [x,w,u]>+<z \vert v><x \vert [y,w,u]>\nn
&~~&+<x \vert w><y \vert [z,u,v]>+<y \vert w><z \vert [x,u,v]>\nn
&~~&+<z \vert w><x \vert [y,u,v]>\Big\}.
\ee
One may equivalently write equation (\ref{triple-def}-iv) in the
form
\be
iv)^\prime&~~&[[x,y,z],u,v]=\big\{\nn
&~~&\alpha \{
         < y \vert v>< z \vert u >-< y \vert u>< z \vert v > \}
         -\beta < u \vert [v,y,z] > \big\}x\nn
&~~&+\alpha\{
         < z \vert v>< x \vert u >-< z \vert u>< x \vert v > \}
         -\beta < u \vert [v,z,x] > \big\}y\nn
&~~&+\alpha\{
         < x \vert v>< y \vert u >-< x \vert u>< y \vert v > \}
         -\beta < u \vert [v,x,y] > \big\}z\nn
&~~&-\beta\big\{
         < x \vert v > [u,y,z]+< y \vert v > [u,z,x]+
         < z \vert v > [u,x,y]\nn
         &~~&
         +< x \vert u > [v,z,y]+< y \vert u > [v,x,z]+
         < z \vert u > [v,y,x]\big\}.
\ee
This triple system is general enough for the specific
application in section \ref{SEC-3}. But, if we would like to consider
physical systems where the coupling constant is not constant
but a function of some parameter $\theta$, as e.g. the energy,
we have to introduce a $\theta$-dependent triple product. Of
special interest is the QCD, where one has the phenomenon of a
so-called ``running coupling constant'', see discussions in
subsections \ref{Sec-3b} and \ref{Sec-3c} below.

The $\theta$-dependent triple system is defined via the
$\theta$-dependent triple product -- note the positions of
variables in LHS and RHS --
\be\label{2-4}
[z,x,y]_\theta&:=&P(\theta )[x,y,z]\nn
&&+A(\theta )< x \vert y> z+B(\theta )< z \vert x> y
  +C(\theta )< z \vert y> x,
\ee
where $A(\theta),B(\theta),C(\theta),P(\theta)$ are functions of
the parameter, see Okubo \cite{Okub2,Okub3}.

\subsection{\label{Sec-2b}Relation of orthogonal triple systems
to the Yang-Baxter equation}

The idea to construct regular vertex functions of non-linear
spinor field models with the help of triple systems arose from
the following connection of orthogonal triple systems and the 
Yang-Baxter--equation (YB), which is known to be a sort of
``integrability condition''. This ``regularization'' will be
used in section \ref{SEC-3}. Now we turn to the formulation of
the Yang-Baxter equation within {\it special}\/ triple systems.
Let $\{e_a\}$ be an arbitrary base of $\V$ and define a dual
base $\{e^b\}$ with help of the scalar product by 
\be
<e^b\vert e_a> &:=& \delta_a^b.
\ee
Strictly spoken, $e^b$ is an {\it image} of a covector in the
vector space $\V$, since we couldn't write down a scalar product
otherwise. Because of the non-degeneracy of the scalar product,
we have an isomorphism between $\V^\ast$ and $\V$ which allows
such a writing. 

Now define the Yang-Baxter--matrix $R^{ab}_{cd}(\theta )$ as
\be
R^{ab}_{cd}(\theta )e_a=[e^b,e_c,e_d]_\theta .
\ee
One is now able to write the Yang-Baxter equation alternatively
in matrix or triple product form \cite{Okub2}. The equivalent
expressions are given with $\theta^\prime=\theta +
\theta^{\prime\prime}$ as either
\be\label{YBE}
R^{b^\prime a^\prime}_{a_1a_2}(\theta )
R^{c^\prime a_2}_{cd}(\theta^\prime )
R^{c_2b_2}_{b^\prime c^\prime}(\theta^{\prime\prime} )
&=&
R^{c^\prime b^\prime}_{b_1c_1}(\theta^{\prime\prime} )
R^{c_2a^\prime}_{a_1c^\prime}(\theta^\prime )
R^{b_2a_2}_{a^\prime b^\prime}(\theta )
\ee
or
\be\label{YBE-T}
[v,[u,e_j,z]_{\theta^\prime},[e^j,x,y]_{\theta}]_{\theta^\pprime}
&=&
[u,[v,e_j,x]_{\theta^\prime},[e^j,z,y]_{\theta^\pprime}]_\theta,
\ee
where we have used the abbreviations $x=e_{a_1}$, $y=e_{b_1}$,
$z=e_{c_1}$, $u=e^{a_2}$ and $v=e^{c_2}$. Usually $\theta$ is
called the ``rapidity parameter''. 

Now it is well known from the theory of integrable systems, that
solutions of this equation behave regular. Hence, a triple
system which additionally satisfy equation (\ref{YBE-T})
provides one with a solution of the Yang-Baxter--equation and is
supposed to provide a good candidate for a vertex function in a
non-linear spinor field model. 

We will see in the next paragraph that triple systems are
very restricted. Nevertheless, it is sufficient for a first
application to study only non-$\theta$-dependent triple
systems, which are of course {\it not}\/ solutions of the
YB-equation. A discussion of the generalization will be given in
the section \ref{Sec-3b}. 

\subsection{\label{Sec-2c}Motivation of the triple system relations}

To give a further motivation of the triple system relation, we
assume $L$ to be a semisimple (simple) Lie--algebra and $\V$ to
be an irreducible $L$-module -- representation space with
irreducible representation.

Any tensor product of such representation spaces can be
decomposed due to the symmetric group. Hence we can decompose
$\V\otimes \V$ into symmetric $(\V\otimes \V)_S$ and
antisymmetric $(\V\otimes \V)_A$ parts as
\be
(\V\otimes \V):=(\V\otimes \V)_S + (\V\otimes \V)_A.
\ee
For higher tensor products one may use Young tableaux to
characterize the appropriate decomposition \cite{Hame62}. This
tableaux can be equivalently described by partitions, e.g.
$[1^3]$ for three totally antisymmetric spaces, $[3]$ for
totally symmetric and $[2,1]$ for a mixed symmetry.

If the $L$-module $\V$ ($\V=[1]$) fulfills the conditions
\be \label{Hom}
i)  &~~& Dim~Hom(({\cal V}\otimes{\cal V})_S\rightarrow {\bf K})=1\nn
ii) &~~& Dim~Hom([1^3]\rightarrow {\cal V})=1\nn
iii)&~~& Dim~Hom([1^4]\rightarrow {\bf K})=1\nn
iv) &~~& Dim~Hom(([1^3]\otimes[1^3])_S\rightarrow {\bf K})\leq 2,
\ee
where $Dim {\cal W}$ is the dimension of the space of
homomorphisms ${\cal W} = Hom({\cal W}_1 \rightarrow {\cal W}_2)$
of $L$-modules from ${\cal W}_1$ into ${\cal W}_2$ which are
compatible with the action of $L$, then (\ref{Hom}) implys the
relations of the triple system (\ref{triple-def}).

(\ref{Hom}-i) provides one the unique existence of a symmetric
inner product. Because of the invariance of $\V$ under the
action of $L$, this inner product is non-degenerate. From this
we can conclude that there exists a unique isomorphism from
$\V^\ast$ onto $\V$ which allows the identification of $\V^\ast$
and $\V$. Because we have restricted ourselves to orthogonal
triple systems, one can conclude from eq. (\ref{triple-def}-i)
the following symmetry of the YB--$R$--matrix:
\be
R^{ab}_{cd}(\theta )
&=&R^{ba}_{dc}(\theta ).
\ee
This is an additional symmetry, but symplectic cases are also
possible. 

The existence of one and only one total antisymmetric triple
product is an outcome of the condition (\ref{Hom}-ii). This
condition restricts the dimension of $\V$.

The restriction (\ref{Hom}-iii) yields the possibility to build
a unique scalar product between an element from $\V$ and a
triple product.

For the interpretation of (\ref{Hom}-iv) we notice, that triple
systems are closely related to algebras with composition 
law. Let
\be
\times &:& \V \times \V \mapsto \V \nn
       && (x;y) \rightarrow x\times y =z,
\ee
then composition means the validity of the following identity
\be
<x\times y \vert x\times y> &=&
<x\vert x><y\vert y>.
\ee
This is very close to the well known formula of determinants
\be
det(AB) = det(A) det(B),
\ee
see \cite{EBBI91}. The connection between (\ref{triple-def}-iv)
and (\ref{Hom}-iv) follows from this structure. First, we have
to notice, that the only (finitely generated) division (${\bf
R}$-)algebras are ${\bf R}$ itself, ${\bf C}$ the complex
numbers, ${\bf H}$ the Hamilton quaternions and ${\bf O}$ the
Cayley octonions, which are of dimension 1,2,4 and 8, see
\cite{EBBI91}. 

To establish the connection, we choose an arbitrary element e,
with the properties
\be
<e \vert e> =1
\ee
and
\be
e\times x = x\times e = x,
\ee
with the above defined binary algebra product. We use the symbol
$\times$ to emphasize the close connection to cross-product
algebras, which can be found in dimension 3 and 7 due to the
reduction by singling out the ``real'' element $e$. With the
associator and the commutator defined as
\be\label{2-17}
i) &~~& (x,y,z) := (x\times y) \times z - x \times ( y \times z) \nn
ii) &~~& [x,y]  := x \times y - y \times x,
\ee
it is possible to express the triple product now as
\be\label{2-18}
{}[x,y,z] &:=&
\frac{1}{2}\Big\{
(x,y,z) + <x\vert e> [y,z] +<y\vert e> [z,x]
+<z\vert e> [x,y] - <z \vert [x,y]> e \Big\}.
\ee
To show the connection to a quaternion algebra (octonion algebra
in 8 dim. case) one can introduce a conjugation
\be\label{2-19}
\bar{x} := 2 <x\vert e>e - x,
\ee
which is obviously an involution. Identifying e with the unit in
${\bf H}$, $e\equiv 1$ and introducing an orthogonal normed base
$\{i,j,k\}$ in the ``imaginary'' part of ${\bf H}$ yields the
analogy. For more complicated triple systems and higher
dimensions see \cite{Okub2}.

\section{\protect\label{Sec-3}Application to a non-linear spinor
filed} 

\subsection{\label{Sec-3a}``Regular vertex'' from a triple system}

Since we are mainly interested in the formulation of a
non-linear spinor field model, we omit the parameter dependent
coupling constant. We explicitly submit that this is an
artificially restriction. One should also note the more
advanced spinor equations of Daviau and Lochak 
\cite{Loch85,Davi91a,Davi93b}, which are much more general. 

Hence we look for spinor theories with a four fermion
interaction like
\be
gV_I^{I_1I_2I_3}\psi_{I_1}\psi_{I_2}\psi_{I_3},
\ee
where $g$ is a real constant, $V_I^{I_1I_2I_3}$ is a constant
vertex and $\psi_{I_1}\psi_{I_2}\psi_{I_3}$ are spinor fields.
This kind of model is found in various very distinct physical
areas, as in solid state theory e.g. the Hubbard model
\cite{Frad} or in the hadronization of QCD as effective low
energy limit, where we would have an Index $I$, which also
includes the color degrees of freedom, e.g. \cite{Rein}. The
important case is, however, the $\theta$-dependent
generalization, which will be discussed in the next subsection.

We omit also the general problems of defining quantized
non-linear spinor field theories, which are discussed elsewhere
\cite{Faus-thesis}, by assuming a sufficient regularization e.g.
by a cutoff method. 

A further problem, arising when quantizing non-linear theories
is ordering. To ``quantize'' a theory means to give a heuristic
concept how to translate a given classical theory into {\it
one}\/ corresponding quantum theory. This is possible in general
for free theories only, where ``$p$'' and ``$q$''--variables
occur only quadratic, see the Groenwald--van Howe theorem
discussed in \cite{Guil90}. The ordering problem was treated in
\cite{Vertex,Faus-thesis} for fermionic fields and will
hereafter not be dealt with. We emphasize, that a treatment
which includes ordering could be given if the current
development would be treated in the algebraic picture promoted
in \cite{Faus-thesis}. 

To be able to determine {\it a particular}\/ vertex function, we
have to search for solutions of the non-$\theta$-dependent
triple system (\ref{triple-def}). Therefore we cite the
following \par 
{\bf theorem:} (Okubo \cite{Okub2})
There are three classes of solutions of the triple system
(\ref{triple-def}): 
\be
a) &~& N=8,\quad\mbox{with~}\alpha^2=\beta\nn
b) &~& N=4,\quad\mbox{with~}\beta=0\nn
c) &~& [[x,y,z],u,v]\equiv0,\quad\alpha=\beta=0.
\ee
Since c) is trivial and a) leads to $L$-modules of dimension 8
and the Cayley octonions, we restrict ourselves to the case b),
simply to exemplify our idea.

Hence we have now $Dim \V=N=4$, and we may choose an orthonormal
base, elements of that kind are called $e_\mu$ where $\mu\in\{1,2,3,4\}$.
Furthermore we obtain in this simple case
\be
<e_\mu\vert e_\nu>&=&\delta_{\mu\nu}.
\ee
This enables us to write (\ref{triple-def}-iii) as
\be
<e_\alpha\vert [e_\mu, e_\nu, e_\lambda]>&=&
\epsilon_{\alpha\mu\nu\lambda},
\ee
where $\epsilon$ is the fully antisymmetric pseudo-tensor of
Levi--Civita. Now, we proceed by distinguishing an element $e$,
which has to obtain the relations  $<e\vert e>=1$, $e\cdot
x=x\cdot e= x$, where the binary dot-multiplication is defined
to be  
\be
\cdot&:&{\cal V}\times{\cal V}\mapsto {\cal V} \nn
&& (x;y) \rightarrow x \cdot y =z.
\ee
Hence, singling out a special element, which provides us also
with an special involution etc., we can connect the triple
product and a bilinear product via associator (\ref{2-17}-i)
and commutator (\ref{2-17}-ii) due to (\ref{2-18}) by
\be
[x,y,z]&=&\frac{1}{2}\Big\{
(x,y,z) + <x\vert e>[y,z]+<y\vert e>[z,x]\nn
&&
+<z\vert e>[x,y]-<z \vert [x,y]>e\Big\}.
\ee
We are very close to quaternions here, consider the conjugation
(\ref{2-19}) we may set $e\equiv 1$ and find an orthogonal base
of the ``imaginary'' part of ${\bf H}$ -- the Hamilton quaterion
algebra -- $\{i,j,k\}$. One obtains then
\be
x\cdot y&=&[x,y,e]\nn
&& +<x\vert e> y+<y\vert e> x-<x\vert y> e,
\ee
which leads for ``imaginary'' -- or say traceless -- elements of
$\V$ to the equations
\be
x \cdot y + y \cdot x &=& -2 <x\vert y> e \\
x \cdot y - y \cdot x &=& 2 [x,y,e],
\ee
which shows the Clifford algebraic nature of the
dot-multiplication, as the close connection of cross-products
and triple products, which can be found in the 7 dimensional
``imaginary'' part of the octonion algebra also \cite{EBBI91}.

An element composed of a real part and an higher dimensional
imaginary ``vector'' part is usually called a para-vector
\cite{Bracks,Sommen}. This concept depends on the grading, where
the imaginary part here is based on the kernel of {\it the}\/
linear form induced by the special element $e$ via,
$\lambda(x):=<x\vert e>$, $\lambda(e)=1$. Since we have
developed the above theory over the complex field, this analogy
is not strict, but could be made so. Nevertheless, we would have
to change the involved concept of gradation.

Now we are prepared to solve our problem for the purpose of
a non-linear spinor field theory. We look for (complex) spinors
of four dimensions, hence elements of a linear space -- spinor
space -- of four dimensions. This spinor space has in a natural
manner an $L$-modul structure due to the action of the
$\gamma$-matrices. The corresponding metric is non degenerate,
symmetric and the spinor representation is irreducible. Hence we
have fulfilled the requirements of the triple system
(\ref{triple-def}) and may apply the above theorem and obtain
its special solution, by setting
\be
V_\alpha^{\alpha_1\alpha_2\alpha_3}&:=&\epsilon_{\alpha\alpha_1
\alpha_2\alpha_3},
\ee
here without color or isospin degrees of freedom. If we develop
the $\epsilon$-tensor into tensor products of $\gamma$ matrices,
we obtain in the symmetric base, which relies on spinor --
charge conjugated spinor spaces the following expression, using
summation convention (s.c.)
\be
\epsilon_{\alpha\alpha_1\alpha_2\alpha_3}&=&\frac{1}{2}\Big\{
\gamma^\mu\gamma^5C\otimes\gamma_\mu\gamma^5C
+C\otimes C
-\gamma^5C\otimes\gamma^5C\Big\}_{\alpha\alpha_1\alpha_2\alpha_3}.
\ee
If Dirac-representation is used, one has $C=i\gamma^2\gamma^0$.
In field theoretic language we have an axial vector, a scalar
and a pseudo scalar part in this expression. If we would like to
have a Fierz-symmetric form, we may use appendix D of
\cite{Grimm} to obtain (s.c.)
\be
\epsilon_{\alpha\alpha_1\alpha_2\alpha_3}&=&\Big\{
-\frac{1}{4}C\otimes C+\frac{3}{4}\gamma^5C\otimes\gamma^5C\nn
&&-\frac{1}{8}\gamma^\mu C\otimes \gamma_\mu C
-\frac{1}{4}\gamma^\mu\gamma^5 C\otimes\gamma_\mu\gamma^5C
+\frac{1}{8}\Sigma_{\mu\nu}C\otimes\Sigma^{\mu\nu}C
\Big\}_{\alpha\alpha_1\alpha_2\alpha_3}.
\ee
In this form, we obtain {\it all}\/ cases of possible couplings,
but with a {\it fixed relative} strength. Hence we are able to
formulate our ``vertex regularized'' non-linear spinor theory,
which reads quantized
\be
(i\gamma^\mu\partial_\mu-m_i)_{\alpha\beta}^{reg}
\Psi_\beta(x)&=&g\epsilon_{\alpha\alpha_1\alpha_2\alpha_3}
\Psi_{\alpha_1}\Psi_{\alpha_2}\Psi_{\alpha_3}(x)\nn
\{\Psi_{\Lambda_1}(t,\vec{\bf r}_1),\Psi^\dagger_{\Lambda_2}(t,
\vec{\bf r}_2)\}_+&=&\delta_{\Lambda_1\Lambda_2}
\delta(\vec{\bf r}_1-\vec{\bf r}_2),
\ee
where {\it reg}\/ means cutoff or something else. We have used
since now a condensed notation, where $\psi_\alpha$ is a spinor
composed of a pair of spinors $\psi_\Lambda$,
$\psi^\dagger_\Lambda$ or respectively the charge conjugated
spinor. 

If we look now at the low energy non-relativistic limit, and
omit also self energy of mass, we obtain
\be
-\frac{\Delta}{2m}\psi_\alpha&=&
g\epsilon_{\alpha\alpha_1\alpha_2\alpha_3}
\psi_{\alpha_1}\psi_{\alpha_2}\psi_{\alpha_3}(x).
\ee
This is the equation of motion of electrons in a
BCS-superconductor after the phonon--electron interaction has
been removed in favor of the local electron--electron
interaction \cite{Fett71}.

The ``regularity'' of this type of vertex function was studied
first time in \cite{Vertex}. A fully quantum filed theoretic
treatment, based on generating functionals was given in
\cite{Faus-thesis}. There and in \cite{Mandel,FDirac} we found a
connection between regularity and the representation and
gradation of the theory. This will be important in the next
subsection. However, we emphasize, that due to the kinetic term,
the theory is, without say a cutoff, still singular.

\subsection{\label{Sec-3b}$\theta$-dependence and ordering}

Since we have restricted ourself to the simple
non-$\theta$-dependent triple system (\ref{triple-def}), we want to
give some further aspects of the general case in this
subsection. 

The relation of $\theta$-dependent and non-$\theta$-dependent
triple systems was given in (\ref{2-4}), where the scalar
functions $A(\theta),B(\theta),C(\theta),P(\theta)$ of the
``rapidity parameter'' $\theta$ were defined also. This triple
system is general and {\it not}\/ subjected to the
YB-equation, which can be seen as an {\it additional}\/
constraint due to (\ref{YBE-T}) of the triple system.

Now, we found in \cite{Vertex} the following relation for
elements of the appropriate $L$-modul
\be\label{normal}
a {\dwedge} b {\dwedge} c &:=&
a \wedge b \wedge c
+F_{ab}c+F_{bc}a+F_{ca}b,
\ee
which was obtained by a normal-ordering procedure. This does
reduce in QFT Fock space to the usual Wick-Dyson normal-ordering
\cite{Vertex}. The wedge $\wedge$ and dotted-wedge $\dwedge$ are
two different representations of the exterior algebra onto the
considered Clifford algebra -- sic. quantized exterior algebra
--, which induce there different gradings \cite{Vertex}. Recall,
that normal-ordering is performed as the lowest step of the
transition of QFT generating functionals to connected ones.
Hence, a normal-ordering extracts the two-particle correlation
and leads to an two-particle irreducible generating functional.
This technique is well known from statistical mechanics
\cite{StumpfRieckers}.  

Since the above defined systems are antisymmetric and noticing
(\ref{triple-def}-iii), we may write
\be
[x,y,z] &\equiv& x \wedge y \wedge z
\ee
for the non-$\theta$-dependent triple product and
\be
[z,x,y]_\theta &\equiv& x {\dwedge} y {\dwedge} z
\ee
for the -- here antisymmetric -- $\theta$-dependent triple
product. If we would define now 
\be
P(\theta) &\equiv& 1 \nn
A(\theta)<x\vert y> &\equiv& F_{xy} \nn
B(\theta)<z\vert x> &\equiv& F_{zx} \nn
C(\theta)<z\vert y> &\equiv& F_{zy},
\ee
we would arrive at the same equation as for the normal-ordering
(\ref{normal}) which was found in \cite{Vertex}. The YB-equation
would thus lead to a further requirement which has to be
fulfilled by the functions $F_{ab}$. In \cite{Fau-pos,Faus-thesis} we
have shown, that $F$ is nothing but the {\it propagator}\/ of
the full, i.e. interacting, theory. It was also shown, that $F$
determines the proper grading of the involved algebras and hence
such notations as ``particle number, scalar, vector, tensor,
$\ldots$'' etc., see also \cite{Mandel,FDirac}.

Looking for the solutions of the YB-equation, found by Okubo for
the above special case, we arrive at
\be
\frac{A(\theta)}{P(\theta)} &=& -a \nn 
\frac{B(\theta)}{P(\theta)} &=& a+b\theta \nn
\frac{C(\theta)}{P(\theta)} &=& a + \frac{a^2-\alpha}{b\theta}
\ee
and hence at
\be\label{3-20}
[z,x,y]_\theta &=&P(\theta)\Big\{
[x,y,z] - a<x \vert y>z +(a+b\theta)<z \vert x>y
        +(a+\frac{a^2-\alpha}{b\theta})<z \vert y>x \Big\}.
\ee
We obtain thus, up to the overall factor $P(\theta)$, a
generalized ``normal-ordering'' procedure due to the
YB-symmetry. The propagator function $F_{xy}$ is no longer
isotrop, but depends on the position of the factors which are
{\it contracted}.\/ Hence, with the above ordering of variables
we have
\be
F^1_{xy} &=& -a<x \vert y> \nn
F^2_{zx} &=& (a+b\theta)<z \vert x> \nn
F^3_{zy} &=& (a+\frac{a^2-\alpha}{b\theta})<z \vert y>.
\ee
We may arrive at the usual isotropic ordering procedure up to the
overall factor $P(\theta)\vert_{\theta=-2a/b}$ by letting
simultaneous 
\be
b\theta &\rightarrow& -2a \nn
\alpha  &\rightarrow& ~5a^2,
\ee
which reduces the three different non-isotropic propagators to
$F_{st}=-a<s\vert t>$ and hence to
\be
[z,x,y]_{-2a/b} &=&P(\frac{-2a}{b})\Big\{
[x,y,z] + F_{xy}z + F_{zx}y + F_{zy}x \Big\}.
\ee
This is up to the overall $P$-factor equivalent to
(\ref{normal}) and proofs thereby the existence of the
appropriate limit. Defining in general (in our case)
\be
<e^a \vert [e^b,e_c,e_d]> &=& C_{ab}^{cd}
\quad\quad\quad\quad\quad\quad (\cong \epsilon_{ab}^{cd}) \nn
<e^a\vert e^b> &=& g^{ab} 
\quad\quad\quad\quad\quad\quad~ (\cong \delta^{ab}) \nn
<e_a\vert e_b> &=& g_{ab}=(g^{ab})^{-1} 
\quad\quad (\cong \delta_{ab}),
\ee
we arrive at the following relation for the YB-$R$-matrix
\be
R_{cd}^{ab}(\theta) &=&
P(\theta) C_{cd}^{ab} + A(\theta)g_{cd}g^{ab} 
+B(\theta)\delta_d^a\delta_c^b
+C(\theta)\delta_c^a\delta_d^b
\ee
respectively
\be
R_{cd}^{ab}(\theta) &=&
P(\theta)\Big\{
\epsilon_{cd}^{ab} - a \delta_{cd}\delta^{ab} 
+(a+b\theta)\delta_d^a\delta_c^b
+(a+\frac{a^2-\alpha}{b\theta})\delta_c^a\delta_d^b\Big\} \nn
( &\cong&
P(\frac{-2a}{b})\Big\{
\epsilon_{cd}^{ab} - a \delta_{cd}\delta^{ab} 
-a\delta_d^a\delta_c^b
-a\delta_c^a\delta_d^b\Big\} ) .
\ee
The overall factor $P(\theta)$ would in our spinor model lead to
a $\theta$-dependent coupling constant $g(\theta) = gP(\theta)$.
But indeed this is a formal analogy and further investigations
on this topic are necessary.

The importance of the obtained result is not founded in the
special model, but in a novel and unexpected relation to recent
effective models in QCD \cite{KL}. These authors considered
(randomly distributed) anisotropic color-interactions discussing
confinement in an effective non-linear spinor field model,
essentially equivalent in structure to our model.

A further important application of such models lays in
high-T$_c$-super conductor models, which can be described by
analogous methods. Both models, in high energy and solid state
physics, rely basically on the so called Anderson localization
\cite{Anderson}, which was the starting point for the model
considered in \cite{KL} and ultimately lead to the anisotropic
interaction there.

In contrast to these considerations, our approach was motivated
from an algebraic setting, which was modeled to obtain somehow
distinguished -- regular -- interactions which are related to
the generalized $q$-symmetry. Some consequences of this results
will be discussed in the next section.

\subsection{\label{Sec-3c}On the nature of quantum deformations}

Since we have connected in a totally different way the
YB-symmetry and hence therewith deformed symmetries with
physical theories, we may try to give some alternative comments
on the physical meaning of $q$-deformed symmetries and
variables. For some expository texts see \cite{Bidenharn}.

Usually, one can develop from the quantum plane $A_q^{p\vert 0}$
and the dual $A_q^{0\vert p}$ the theory of deformations
\cite{Manin}. One is then confronted with the intriguing concept
of an {\it non-commutative point space}. There are very much
speculations on the physical meaning of such constructions. Most
of this considerations do connect the discrete structure, which
comes along with $q$-deformation, with the physics at the Planck
scale \cite{Majid}. Since we are interested in low energy solid
state physics or the low energy behavior of QCD, we cannot
account for the Planck scale and related ideas of a discrete
space-time \cite{Mueller-H}. Additionally, one arrives at
$q$-deformed structures by studying differential calculi on
Lattices \cite{Mueller-HDimakis}. This lead already to
speculations, that a discrete space-time may be found underlying
conventional continuous space-time. Such considerations may be
the origin of the extensive study of deformed Lorentz and
Poicare symmetry \cite{Wess+Co}. The difficult point is, that
the deformation parameters, if there are many, are {\it not}\/
subjected to the dynamics of the system. Deformation provides
the background of a theory, but not the details. From QFT it is
clear, that the dynamics of a system {\it has to choose}\/ a
particular -- non-Fock, if the theory is not free --
representation. The representation is in general subjected to
the dynamics, which forbids in general a fixed, non-dynamical,
deformation. 

Our treatment of the four fermion local interaction suggests
another attempt to explain this structure. The most important
fact is the usability of the generalized symmetry. This is the
foundation of the integrability of systems which satisfy the
YB-equation. If we reject the quantum plane as a {\it point
space},\/ we remain with all benefits of the involved symmetry. 

We state therefore, that the $q$-variables model {\it some}\/
aspects of composites in multi-particle systems. Non-commutativ
spaces therefore does not conflict with the usual physical point
space. The preparation of composites has to be conserved {\it
somehow}\/ under the dynamics. This cannot be an ordinary
isometry, as long as the composite is not entirely rigid and the
composite density is low -- weak interaction of composites assumed.

Our simple consideration above supplies this point of view. The
local electron-electron interaction arose due to the elimination
of the intermediate phonon interaction. The same situation is
considered in effective QCD theories \cite{KL}. Hence, we may
look at the spinors $\psi_\alpha$ as composed of an electron
(quark) and somehow phonons (gluons), which could be described
by higher spin-tensors. The YB-equation is then on the one hand an
integrability condition and on the other the condition, how to
be able to map the general structure of composites onto another.

We may state, that $q$-deformed symmetry is the symmetry of
composites -- entities with dynamical internal structure. The
$q$-parameters are parameters of this structures, which define
the possible types of ``congruent'' structures, and are
therefore not dynamic. Due to the connection to the usually
employed normal-ordering of QFT, we obtain a direct connection
between $q$-parameters and the {\it state space}\/ of the theory
\cite{Faus-thesis}. The connection of the YB-solutions and
non-isotropic contractions -- sic. propagators -- was outlined
above, see also \cite{Vertex,Mandel,Faus-thesis,FDirac}. Since
contractions rely on two-particle issues, this is a direct way
to see the multi-particle character of $q$-symmetry. An
extensive discussion of the connection of quantum mechanical
state space and $q$-symmetry will be given elsewhere. 

\section{\protect\label{Sec-4}Conclusion}

We developed, by applying triple systems, a method to obtain a
``regular vertex'' in a non-linear spinor field model. However,
the regularity treated in \cite{Vertex,Faus-thesis} was not made
explicit. The main idea was to show the usefulness of triple
products in the determination of distinguished vertex functions.
The easy and elegant handling of YB-symmetry by special triple
systems was used to give a distinguished vertex, which proved to
be in the non-relativistic case the BCS interaction, when phonons
are eliminated in favor of a local electron-electron
interaction. The assumptions underlying the employed triple
system were physically motivated. The requirements are in a
non-mathematical language, that first one assumes a
non-degenerate symmetric scalar product. Second, one forces the
triple product to behave as an {\it effective single field}.\/
Third, there is a scalar product between single fields and
triples. Fourth, there is a scalar product on these effective
fields. 

All in all, these assumptions guarantee the theory to be
considered as an effective theory. This physical background
remains true if $\theta$-dependent triple systems are
considered. 

Furthermore, we gave an outline, how this procedure can be
generalized to the important class of models with parameter
dependent -- running -- coupling constants. Only in this case
one obtains a YB-symmetry. 

The recent interest in anisotropic local four fermion
interactions gives direct relevance to our investigations. Every
solution of the YB-equation involved in the $\theta$-dependent
triple system leads, up to a $\theta$-dependent overall factor,
to a fixed anisotropic vertex function and constrains the
physical possibilities drastically.

The $\theta$-dependent triple system also lead to a generalized
{\it non-isotropic ordering procedure} which was shown to be
equivalent to the usual Wick-Dyson normal-ordering in the
non-$\theta$-dependent limit.

The new approach to deformed symmetries was used to give an
alternative interpretation of $q$-symmetries and $q$-variables.
We rejected $q$-variables as {\it point space}, but looked at
them as a sort of effective coordinates of composites which
exhibit non-trivial internal dynamics. The investigation with
help of triple systems and their relation to generalized
non-isotropic orderings does support this point of view.

\end{document}